\documentclass[%
 reprint,
superscriptaddress,
preprintnumbers,
nofootinbib,
 amsmath,amssymb,
 aps,
]{revtex4-2}

\usepackage{graphicx}
\usepackage{dcolumn}
\usepackage{bm}
\usepackage{hyperref}
\usepackage{color}
\usepackage{float,subcaption}

\hypersetup{
    unicode=false,          
    pdftoolbar=true,        
    pdfmenubar=true,        
    pdffitwindow=false,     
    pdfstartview={FitH},    
    pdftitle={},    
    pdfauthor={},     
    pdfsubject={},   
    pdfcreator={},   
    pdfproducer={}, 
    pdfkeywords={} {} {}, 
    pdfnewwindow=true,      
    colorlinks=true,       
    linkcolor=blue, 
    citecolor=blue,        
    filecolor=magenta,      
    urlcolor=blue,           
	breaklinks=true
} 

\usepackage{amssymb}
\usepackage{amsmath}
\usepackage{amsfonts}
\usepackage{xspace}
\usepackage{bm,bbm}
\usepackage{relsize}
\usepackage{siunitx}
\usepackage{xcolor}
\usepackage{upgreek}

\makeatletter
\DeclareFontFamily{OMX}{MnSymbolE}{}
\DeclareSymbolFont{MnLargeSymbols}{OMX}{MnSymbolE}{m}{n}
\SetSymbolFont{MnLargeSymbols}{bold}{OMX}{MnSymbolE}{b}{n}
\DeclareFontShape{OMX}{MnSymbolE}{m}{n}{
    <-6>  MnSymbolE5
   <6-7>  MnSymbolE6
   <7-8>  MnSymbolE7
   <8-9>  MnSymbolE8
   <9-10> MnSymbolE9
  <10-12> MnSymbolE10
  <12->   MnSymbolE12
}{}
\DeclareFontShape{OMX}{MnSymbolE}{b}{n}{
    <-6>  MnSymbolE-Bold5
   <6-7>  MnSymbolE-Bold6
   <7-8>  MnSymbolE-Bold7
   <8-9>  MnSymbolE-Bold8
   <9-10> MnSymbolE-Bold9
  <10-12> MnSymbolE-Bold10
  <12->   MnSymbolE-Bold12
}{}

\let\llangle\@undefined
\let\rrangle\@undefined
\DeclareMathDelimiter{\llangle}{\mathopen}%
                     {MnLargeSymbols}{'164}{MnLargeSymbols}{'164}
\DeclareMathDelimiter{\rrangle}{\mathclose}%
                     {MnLargeSymbols}{'171}{MnLargeSymbols}{'171}
\makeatother

\let\originalleft\left
\let\originalright\right
\renewcommand{\left}{\mathopen{}\mathclose\bgroup\originalleft}
\renewcommand{\right}{\aftergroup\egroup\originalright}

\usepackage{xr}
\makeatletter

\newcommand*{\addFileDependency}[1]{
\typeout{(#1)}
\@addtofilelist{#1}
\IfFileExists{#1}{}{\typeout{No file #1.}}
}\makeatother

\newtheorem{thm}{Theorem}[section]
\newtheorem{lem}[thm]{Lemma}
\newtheorem{cor}[thm]{Corollary}

\begin{document}

\title{Efficient local classification of parity-based material topology}

\author{Stephan Wong}
\affiliation{Center for Integrated Nanotechnologies, Sandia National Laboratories, Albuquerque, New Mexico 87185, USA}

\author{Ichitaro Yamazaki}
\affiliation{Center for Computing Research, Sandia National Laboratories, Albuquerque, New Mexico 87185, USA}

\author{Chris Siefert}
\affiliation{Center for Computing Research, Sandia National Laboratories, Albuquerque, New Mexico 87185, USA}

\author{Iain Duff}
\affiliation{Computational Mathematics Group, Scientific Computing Department, Rutherford Appleton Laboratory, Oxfordshire, England}

\author{Terry A. Loring}
\affiliation{Department of Mathematics and Statistics, University of New Mexico, Albuquerque, New Mexico, 87131, USA}

\author{Alexander Cerjan}
\affiliation{Center for Integrated Nanotechnologies, Sandia National Laboratories, Albuquerque, New Mexico 87185, USA}

\date{\today}

\begin{abstract}
Although the classification of crystalline materials can be generally handled by momentum-space-based approaches, topological classification of aperiodic materials remains an outstanding challenge, as the absence of translational symmetry renders such conventional approaches inapplicable. Here, we present a numerically efficient real-space framework for classifying parity-based $\mathbb{Z}_2$ topology in aperiodic systems based on the spectral localizer framework and the direct computation of the sign of a Pfaffian associated with a large sparse skew-symmetric matrix. Unlike projector-based or momentum-space-based approaches, our method does not rely on translational symmetry, spectral gaps in the Hamiltonian's bulk, or gapped auxiliary operators such as spin projections, and instead provides a local, energy-resolved topological invariant accompanied by an intrinsic measure of topological protection. A central contribution of this work is the development of a scalable sparse factorization algorithm that enables the reliable determination of the Pfaffian’s sign for large sparse matrices, making the approach practical to realistic physical materials. We apply this framework to identify the quantum spin Hall effect in quasicrystalline class AII systems, including gapless heterostructures, and to diagnose fragile topology in a large $C_2 \mathcal{T}$-symmetric photonic quasicrystal. Overall, our results demonstrate that the spectral localizer, combined with efficient sparse numerical methods, provides a unified and robust tool for diagnosing parity-based topological phases in aperiodic electronic, photonic, and acoustic materials where conventional band-theoretic indexes are inapplicable.
\end{abstract}

\maketitle


\section{Introduction}

Topological insulators are a class of materials characterized by non-trivial topological invariants and are guaranteed through bulk-boundary correspondence to exhibit robust boundary states protected against perturbations~\cite{Chiu2016, Ozawa2019, Ma2019, Cooper2019}.
This robustness is the key to enabling new technological domains, such as spintronics, where spin-momentum locking may be useful for low-power, high-speed electronics \cite{he_topological_2022}, as well as fault-tolerant routes to quantum computing through topological superconductors \cite{qi_topological_2011}. Topological insulators can also serve as precursors to correlated phases at low temperatures when electron interactions become strong \cite{hohenadler_correlation_2013}, revealing novel forms of quantum order.
In crystalline systems, topological classification is traditionally performed using approaches rooted in band theory, where the system's Bloch wavefunctions are used to define invariants such as Chern numbers~\cite{Thouless1982}, the $\mathbb{Z}_2$ index~\cite{Kane2005,fu_topological_2007}, winding numbers~\cite{Asboth2016}, and multipole moments~\cite{Benalcazar2017}.
Modern band-theoretic classification frameworks based on elementary band representations can be applied across databases of crystalline materials, resulting in the prediction of a wide variety of topological phases \cite{bradlyn_topological_2017, vergniory_complete_2019, elcoro_magnetic_2021}.

In contrast, the investigation of topology in non-periodic systems remains challenging, as the lack of translational symmetry renders momentum-space classification formulations inapplicable. 
To address this need, real-space approaches to material classification have been developed, either based on projected position operators such as the Bott index~\cite{Hastings2011}, the Kitaev~\cite{Kitaev2006} or the Bianco-Resta markers~\cite{Bianco2011}, or rooted in the spectral localizer~\cite{Loring2015, Loring2017, Loring2020, Cerjan2024a}.
These theories of real-space markers have been successfully applied to identify Chern- and Weyl-based topology in quasicrystalline~\cite{Tran2015, Bandres2016, Fulga2016, GrossieFonseca2023, Wong2024}, amorphous~\cite{Mitchell2018, Zhou2020, Mitchell2021, Corbae2023}, and disordered~\cite{Sahlberg2023, Lee2025} systems. 
For non-periodic $\mathbb{Z}_2$ topological insulators in class AII of the Altland-Zirnbauer (AZ) scheme~\cite{Altland1997, Schnyder2008, Kitaev2009, Ryu2010} that exhibit the quantum spin Hall effect (QSHE)~\cite{Kane2005}, specific real-space methods have also been developed, such as approaches based on twisted boundary conditions~\cite{Essin2007, Prodan2011, Leung2012}, scattering matrices~\cite{Fulga2011, Fulga2012, Sbierski2014}, non-commutative geometry~\cite{Prodan2010, Prodan2011a}, and the spin Bott index~\cite{Huang2018, Huang2018a, Chen2019, Peng2024}. 
However, these methods for aperiodic class AII systems face significant limitations. 
In particular, these existing local markers assume the presence of a spectral gap in the Hamiltonian and sometimes also a gap in a spin projection, and thus fail in gapless systems that commonly arise in realistic models of disordered or aperiodic topological phases. Moreover, these methods often lack an independent measure of topological protection, i.e., beyond any spectral gap, making it difficult to quantify the robustness of any edge states.
Computationally, the reliance on projectors in these approaches also forces the consideration of dense matrices, limiting these frameworks to low-energy approximations of a material and thus requiring that such a low-energy approximation first be found.
Yet, aperiodic materials represent a crucial outstanding class of materials for broad topological classification that are becoming even more relevant with the advent of twisted van der Waals systems that can exhibit a variety of topological and correlated phases, and in which incommensurate twist angles yield quasiperiodic systems \cite{nuckolls_microscopic_2024}.

Here, we develop a numerically efficient approach rooted in the spectral localizer framework to classify $\mathbb{Z}_2$ parity-based topology across a range of different symmetry classes and demonstrate its application to realistic aperiodic systems without needing a low-energy approximation.
As this framework does not rely on projecting into an occupied subspace, our approach can identify the quantum spin Hall effect in aperiodic class AII systems without requiring assumptions about whether the material exhibits a spectral gap or sufficiently separated spin sectors.
In particular, the material's local topology is diagnosed through the sign of the Pfaffian of a skew-symmetric matrix derived from the spectral localizer, and a key aspect of this study is the release of an efficient scalable algorithm for directly computing a Pfaffian's sign for arbitrary sparse skew-symmetric matrices \cite{wong_sparse_sign_pfaffian_2026}.
In other words, our algorithm can handle materials and heterostructures in arbitrary dimensions described by wave equations, not just tight-binding models.
We illustrate the utility of our approach by both locally classifying the QSHE in a tight-binding quasicrystalline heterostructure and identifying the fragile topology of a $C_2 \mathcal{T}$-symmetric 2D photonic quasicrystal.
More broadly, our numerical approach allows for the efficient computation of $\mathbb{Z}_2$ parity-based invariants in many other AZ classes~\cite{Loring2015}, both enabling the prediction of boundary-localized protected states in natural, photonic~\cite{Ozawa2019}, and acoustic materials~\cite{Ma2019}, as well as identifying candidate materials that may exhibit novel correlated phases.
Moreover, while our algorithm does not directly compute the Pfaffian of a sparse skew-symmetric matrix to find its sign, it can be combined with algorithms to find a skew-symmetric matrix's determinant to yield the Pfaffian, as $\textrm{Pf}(M)= \textrm{sign}(\textrm{Pf}(M)) \sqrt{\det (M)}$.


\section{Spectral localizer \label{sec:2}}

To motivate the development of an efficient numerical algorithm for computing the Pfaffian's sign of a sparse skew-symmetric matrix, we consider the classification of $\mathbb{Z}_2$ topology in non-periodic class AII systems using the spectral localizer framework~\cite{Loring2015}. 
The key advantage of this framework is that it uses an operator-based methodology that incorporates the system's spatial position operators while not relying directly on the computation of a specific eigensubspace.
In particular, the spectral localizer diagnoses the system's local topology from the real-space perspective of atomic limits by assessing whether a system's Hamiltonian $H$ can be path-continued to be commuting with system's positions operators $X,Y$ while preserving the relevant symmetry(s) and a spectral gap of a matrix derived from $H$, $X$, and $Y$.  
As such, the spectral localizer produces an effective method for classifying the system's local topology without spin resolution or band gap assumptions and comes equipped with an inherent measure of protection~\cite{Cerjan2022, Cerjan2022a, Dixon2023, Wong2024, Cerjan2024, Cerjan2024a}.

Consider a system with fermionic time-reversal symmetry $\mathcal{T} = (I \otimes i \sigma_y) \mathcal{K}$, with $\mathcal{K}$ being the complex conjugate operator and $I$ the identity.
By definition, $H$ is time-reversal symmetric, meaning the Hamiltonian can be written in block form as~\cite{Loring2012}
\begin{equation}
\label{eq:H_tri}
H = 
\left(
\begin{array}{cc}
A & B \\
-B^* & A^*
\end{array}
\right),
\end{equation}
where $^*$ stands for complex conjugation. 
Physically, this block form reveals the nature of the QSHE, which relies on the particular couplings contained in $B$ between some internal degrees of freedom in $A$, e.g., electron spins. Likewise, the position operators also possess a similar structure as at least two degrees of freedom exist at every site.
A spectral localizer is then generated by pairing a system's Hamiltonian and position operators each with a different element of a sufficiently large irreducible Clifford representation. For two-dimensional systems where there are only three such operators, $H$, $X$, and $Y$, the Pauli matrices can be used as the Clifford representation, such that
\begin{multline} \label{eq:localizer}
   L_{(\mathbf{x},E)}(X, Y, H) = (H-EI) \otimes \sigma_z  \\
   + \kappa (X-xI) \otimes \sigma_x + \kappa (Y-yI) \otimes \sigma_y
\end{multline}
is a spectral localizer and by construction is Hermitian, though note that any choice of pairing between the system's operators and the Pauli matrices will work. Here, the choices of spatial location $\mathbf{x} = (x,y)$ and energy $E$ determine where in position-energy space the system's topology is being diagnosed, and $\kappa > 0$ is a hyperparameter used to make the units consistent between the Hamiltonian and position operators, as well as to balance the spectral weight between position and energy~\cite{Cerjan2024a}. 

In general, the philosophy of the spectral localizer framework is to translate physical symmetries into a specific matrix structure of $L_{(\mathbf{x},E)}$ to reveal a topological invariant rooted in matrix homotopy. By itself, the fermionic time-reversal symmetry of $H$, $X$, and $Y$ is insufficient to guarantee a useful structure of $L_{(\mathbf{x},E)}$; on the spectral localizer, this symmetry manifests as
\begin{equation}
(\mathcal{T}\otimes I) L_{(\mathbf{x},E)}(\mathcal{T}\otimes I)^{-1} = L_{(\mathbf{x},E)}. \label{eq:LTsym}
\end{equation}
Here, there is a slight abuse of notation, as the $\mathcal{K}$ operation in $\mathcal{T}$ is only applied to the material operators $H,X,Y$ and not on the Pauli matrices.
However, note that the incorporation of the Pauli matrices introduces an additional symmetry to $L_{(\mathbf{x},E)}$ as the Pauli matrices are all odd with respect to an effective fermionic time-reversal symmetry $\mathcal{T}_\textrm{P} = i\sigma_y \mathcal{K}$, such that $\mathcal{T}_\textrm{P} \sigma_j \mathcal{T}_\textrm{P}^{-1} = - \sigma_j$ and 
\begin{equation}
    (I \otimes \mathcal{T}_\textrm{P}) L_{(\mathbf{x},E)} (I \otimes \mathcal{T}_\textrm{P})^{-1} = -L_{(\mathbf{x},E)}, \label{eq:LTPsym}
\end{equation}
leveraging a similar notational abuse with $\mathcal{K}$ in $\mathcal{T}_\textrm{P}$ only applying to the Pauli matrices.
Thus, the existence of both of these individual-sector symmetries means that the spectral localizer for 2D class AII systems possesses a third symmetry, 
\begin{equation}
(\mathcal{T} \otimes \mathcal{T}_\textrm{P}) L_{(\mathbf{x},E)} (\mathcal{T} \otimes \mathcal{T}_\textrm{P})^{-1} = -L_{(\mathbf{x},E)}, \label{eq:LCsym}
\end{equation}
which resembles an effective charge conjugation symmetry on the spectral localizer with $(\mathcal{T} \otimes \mathcal{T}_\textrm{P})^2 = I$. Altogether, this means that the spectral localizer for 2D class AII systems can be written in a basis such that $Q^\dagger L_{(\mathbf{x},E)} Q$ is purely imaginary,
\begin{equation}
    (Q^\dagger L_{(\mathbf{x},E)} Q)^\top = - Q^\dagger L_{(\mathbf{x},E)} Q,
\end{equation}
while remaining Hermitian and $Q$ unitary. This result is analogous to how the Hamiltonians of 0D class D systems can also always be written in a basis where the Hamiltonian is purely imaginary.

The existence of a basis in which a 2D class AII spectral localizer is purely imaginary, and thus skew-symmetric, means that $L_{(\mathbf{x},E)}$ exists in one of two different homotopy classes, distinguished by~\cite{Loring2015}
\begin{equation}
\label{eq:local_index}
\xi^\textrm{L}_{(\mathbf{x},E)}(X, Y, H) = 
\textrm{sign} \left[ \textrm{Pf}\left( Q^\dagger L_{(\mathbf{x},E)}(X, Y, H) Q \right) \right] \in \mathbb{Z}_2,
\end{equation}
so long as the spectral localizer remains gapped, i.e., none of its eigenvalues are $0$.
Here, $\text{sign}(\alpha)$ is the sign of $\alpha$ and $\textrm{Pf}(M)$ is the Pfaffian of a skew-symmetric matrix. The key point is that perturbations to $L_{(\mathbf{x},E)}$, either in the form of perturbing the system or shifting the choice of $(\mathbf{x},E)$, must possess the same invariant as that of the unperturbed $\xi^\textrm{L}_{(\mathbf{x},E)}(X, Y, H)$ so long as the spectral gap of $L_{(\mathbf{x},E)}$ at $0$ remains open under the perturbation, as a matrix's Pfaffian cannot change sign without two of its eigenvalues reaching $0$. Moreover, as $L_{(\mathbf{x},E)}$ is Hermitian, the movement of its eigenvalues under such perturbations is limited by Weyl's inequality \cite{weyl_asymptotische_1912,Bhatia1997}. Thus, the local index $\xi^\textrm{L}_{(\mathbf{x},E)}$ is accompanied by a measure of robustness given by the local gap
\begin{equation}  
\label{eq:local_gap}
\mu_{(\mathbf{x},E)}(\mathbf{X}, H) = \\ 
\min \big[ | {\rm Spec} \left( L_{(\mathbf{x},E)}(\mathbf{X}, H) \right) | \big]
,
\end{equation}
where $\text{Spec}[M]$ is the spectrum of $M$. In addition, locations where $\mu_{(\mathbf{x},E)} \approx 0$ indicate the existence of an approximately localized state $\boldsymbol{\upphi}$ in $H$, with $H \boldsymbol{\upphi} \approx E \boldsymbol{\upphi}$, $X \boldsymbol{\upphi} \approx x \boldsymbol{\upphi}$, and $Y \boldsymbol{\upphi} \approx y \boldsymbol{\upphi}$ \cite{Cerjan2023a}. 

Overall, this analysis shows that the topology of a 2D class AII system at a given location in position-energy space $(\mathbf{x},E)$ can be locally classified by $\xi^\textrm{L}_{(\mathbf{x},E)}$, which carries with it a measure of protection $\mu_{(\mathbf{x},E)}$, and at locations where the topology changes and $\mu_{(\mathbf{x},E)} \rightarrow 0$, the system is exhibits an approximately localized state, yielding a bulk-boundary correspondence. By considering an atomic limit of a gapped Hamiltonian, one can conclude that $\xi^\text{L}_{(\mathbf{x},E)} = 1$ corresponds to a location in position-energy space where the system is trivial, and $\xi^\text{L}_{(\mathbf{x},E)} = -1$ denotes the topological phase. In addition, one can find that the unitary $Q$ that transforms $L_{(\mathbf{x},E)}$ in Eq.~\eqref{eq:localizer} to be purely imaginary is
\begin{equation}
    Q = \frac{1}{\sqrt{2}} \left[ I \otimes I - i (I \otimes i \sigma_y) \otimes (i \sigma_y) \right].
\end{equation}
Altogether, unlike other real-space approaches to $\mathbb{Z}_2$ topology~\cite{Essin2007, Prodan2011, Fulga2011, Prodan2010, Huang2018a}, the identification of the real-space $\mathbb{Z}_2$ topological phase within the spectral localizer framework can be computed without the need of bulk band gap or other ``gapped subspaces'', such as for the invariant relying on the spin projectors for example.


\section{Computation of the sign of a Pfaffian}

The ability to classify parity-based material topology and identify the associated boundary-localized states using the sign of a Pfaffian of a large sparse matrix has the potential to unlock the rapid classification of a wide range of aperiodic materials. However, realizing this promise requires a numerical implementation that can take advantage of the sparsity of $L_{(\mathbf{x},E)}$, as $H$ and $\mathbf{X}$ are typically sparse. Generally, there are several choices that must be made to compute the sign of the Pfaffian, including conventions used to define the Pfaffian, and as such, it is easy to numerically compute the sign of the Pfaffian incorrectly or inefficiently. Instead, we seek a structure that is amenable to efficient numerical computation of the Pfaffian's sign without requiring the calculation of the full Pfaffian, as this is the homotopy invariant of skew-symmetric matrices. 

\subsection{Mathematics}

Fundamentally, the Pfaffian of a skew-symmetric matrix can be understood as function that is a polynomial in the elements of the matrix, and which defines a continuous square root of the matrix's determinant. The Pfaffian's role in distinguishing two distinct homotopy classes of skew-symmetric matrices can be seen by considering the two matrices
\begin{equation}
    S_+ = \left[ \begin{array}{cc}
        0 & \alpha \\ -\alpha & 0
    \end{array} \right], \qquad
    S_- = \left[ \begin{array}{cc}
        0 & -\alpha \\ \alpha & 0
    \end{array} \right].
\end{equation}
While $\det(S_+) = \det(S_-) = \alpha^2$, neither matrix can be deformed into the other without the path in $\alpha$ either becoming non-invertible when $\alpha \rightarrow 0$ or the path becoming non-skew-symmetric. Thus, $S_\pm$ are elements in distinct homotopy classes of invertible matrices, distinguished by the sign of the element in the upper right corner, i.e., by $\textrm{sign}[\textrm{Pf}(S_\pm)]$. More generally, for a $2n$-by-$2n$ skew-symmetric tridiagonal matrix
\begin{equation}
T=\left[\begin{array}{cccccc}
0 & t_{1}\\
-t_{1} & 0 & t_{2}\\
 & -t_{2} & 0 & t_{3}\\
 &  & -t_{3} & 0 & \ddots\\
 &  &  & \ddots & \ddots & t_{2n-1}\\
 &  &  &  & -t_{2n-1} & 0
\end{array}\right],
\end{equation}
the Pfaffian is given by a product based on the elements from the upper diagonal
\begin{equation}
\textrm{Pf}(T)=\prod_{j=1}^{n}t_{2j-1}. \label{eq:pfT}
\end{equation}
Thus, for a tridiagonal skew-symmetric matrix, $\textrm{sign}[\textrm{Pf}(T)]$ can be calculated by simply looking at the signs of each $t_{2j-1}$, without carrying out the full multiplication.

To use Eq.~\eqref{eq:pfT} to achieve an efficient numerical approach for a general skew-symmetric matrix, two further results are required. First, any skew-symmetric matrix $S$ can be factorized using a lower-triangular matrix $L$ to be tridiagonal, such that
\begin{equation}
    S = L T L^\top, \label{eq:factor}
\end{equation}
see Supplementary Material Lemma I.4. Second, that the Pfaffian of such a factorization obeys
\begin{equation}
    \textrm{Pf}(L T L^\top) = \det(L) \textrm{Pf}(T),
\end{equation}
see Supplementary Material Theorem I.10. Moreover, as $L$ is lower triangular, $\det(L)$ can be found by simply taking the product of the diagonal entries in $L$. Altogether then,
\begin{equation}
    \textrm{sign}[\textrm{Pf}(S)] = \left( \prod_{j=1}^{2n} \textrm{sign}(L_{j,j}) \right) \left( \prod_{j=1}^{n} \textrm{sign}(T_{2j-1,2j-1}) \right),
\end{equation}
yielding an approach that has the possibility of realizing an efficient computational method so long as a fast algorithm for performing the factorization Eq.~\eqref{eq:factor} can take advantage of any sparseness in $S$.

\subsection{Computation \label{sec:3b}}

To implement a fast, sparse skew-symmetric factorization method, we leveraged the software package Tacho~\cite{Tacho,ShyLU}, available as part of the open-source software framework Trilinos~\cite{trilinos}, and took inspiration for the permutation strategies from those originally developed by Duff~\cite{duff2009design}. Tacho's routines are based on multifrontal factorization while exploiting the dense \textit{supernodal} blocks that appear during the factorization process.
Here, we have extended the serial factorization routine of Tacho to compute the $LTL^\top$ factorization of a skew-symmetric matrix, 
\[
   P S P^\top = LTL^\top, 
\]
where $P$ is a row permutation matrix, $T$ is a skew-symmetric 2-by-2 block diagonal matrix, and $L$ is a unit-diagonal lower-triangular matrix.

To maintain numerical stability of the skew-symmetric factorization,
we combined two types of permutation strategies,
\[
   P = P_n P_s,
\]
where $P_s$ and $P_n$ represent the row permutations applied 
before and during the numerical factorization, respectively.
As Tacho is primarily designed to obtain high-performance on a GPU, the pivoting $P_n$ is applied only within the supernodal blocks. Hence, the purpose of the initial permutation $P_s$ is twofold, to both enhance the stability and efficiency of the factorization.

To enhance the numerical stability of the factorization, 
we first permute the rows and columns of $S$ to move the nonzero entries with large numerical values to the subdiagonal, such that it is likely for $T$ to have stable 2-by-2 pivots.
We compute this initial permutation, $P_{s_1}$, based on the maximum cardinality or weighted matching algorithm~\cite{duff1998}. 

\begin{figure*}[t]
\center
\includegraphics[width=2\columnwidth]{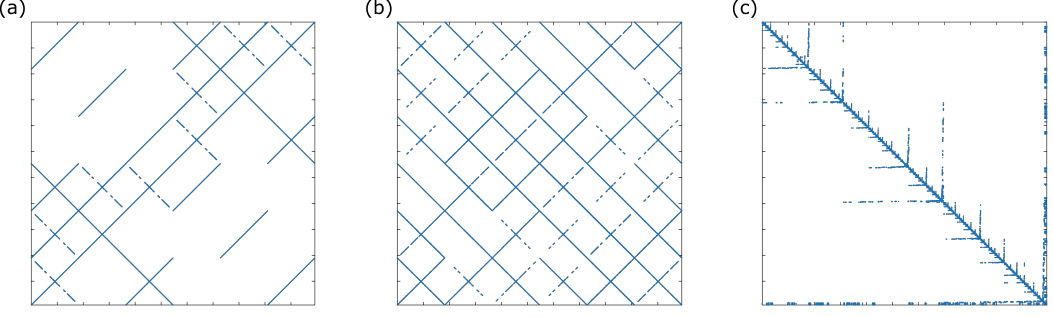}
\caption{
\textbf{Sparsity structure during the initial skew-symmetric permutations.}
Structure of the non-zero elements of the original matrix $S$ (a), after matching $P_{s_1} S P_{s_1}^\top$ (b), and after fill-reducing reordering $P_{s} S P_{s}^\top$ (c) where $S$ is a spectral localizer in a skew-symmetric basis for a 2D photonic quasicrystal. For this example, $S$ is $546016$-by-$546016$ and has $3244872$ nonzero elements. \label{fig:matrix}}
\end{figure*}

Second, we permute the rows and columns of $P_{s_1} S P_{s_1}^\top$ to enhance the performance of the numerical factorization. In our numerical experiments,
we use the nested dissection algorithm implemented in METIS~\cite{metis} to compute the second permutation, $P_{s_2}$.
This matrix reordering strategy is designed to reduce the number of new nonzero entries in $L$, so-called \emph{fills}, introduced during the factorization,
lowering both the storage and computational costs of the factorization.
Before computing $P_{s_2}$, we compress the matrix $S_1 = P_{s_1} S P_{s_1}^\top$ with the 2-by-2 blocks such that the dimension of the resulting matrix $\widetilde{S}_1$ is half of the dimension of $S_1$ and the $(i,j)$th entry of $\widetilde{S}_1$ is nonzero only if the corresponding 2-by-2 block of $S_1$ is non-empty. 
We then compute $P_{s_2}$ based on the fill-reducing reordering permutation $\widetilde{P}_{s_1}$ of the compressed matrix $\widetilde{S}_1$. In this way, the candidates for the 2-by-2
pivots computed by the maximum-weighted matching stay together in $P_{s_2} (P_{s_1} S P_{s_1}^\top) P_{s_2}^\top$. The sparsity structure of an example skew-symmetric matrix before and after each step of the initial permutation process is shown in Fig.~\ref{fig:matrix}. 

\begin{figure}[t]
\center
\includegraphics[width=\columnwidth]{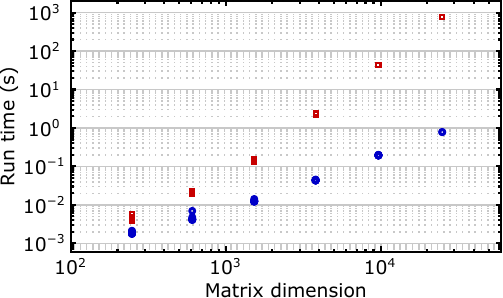}
\caption{
\textbf{Runtime trend comparison of sparse and dense approaches.}
Runtime to calculate $\textrm{sign}[\textrm{Pf}(S)]$ as a function of matrix dimension using the sparse skew-symmetric factorization (blue circles) compared against a dense approach based on the Hessenberg factorization of $S$. 
Here, $S$ is the spectral localizer in a skew-symmetric basis of a 2D tight-binding lattice formed from a Penrose tiling--based quasicrystal, such that the matrix dimension reflects the size of the tight-binding lattice. Different dots at the same matrix size are for different coupling strengths between the sites in the quasicrystal lattice.
\label{fig:timing}}
\end{figure}

Altogether, our approach yields an efficient algorithm to compute the Pfaffian's sign for generic sparse skew-symmetric matrices using
\begin{equation}
    \textrm{sign}[\textrm{Pf}(S)] = (\textrm{sign}[\det(L)])(\textrm{sign}[\det(P)])\textrm{sign}[\textrm{Pf}(T)].
\end{equation}
Moreover, as expected, our implementation achieves improved runtime scaling as a function of matrix size compared against dense algorithms, see Fig.~\ref{fig:timing}.

Before concluding this section, we note that if $S$ is both skew-symmetric and Hermitian, it is purely imaginary; for example, all of the entries of $Q^\dagger L_{(\mathbf{x},E)} Q$ from Sec.~\ref{sec:2} are imaginary. For such cases, $\pm i$ should be inserted to make $S$ real while preserving its skew-symmetry and so as to enable the use of algorithms tailored for real matrices that tend to be more efficient. As $\textrm{Pf}(cS) = c^n \textrm{Pf}(S)$ for an $S$ of size $2n$-by-$2n$, the insertion of $\pm i$ has a predictable effect on $\textrm{sign}[\textrm{Pf}(S)]$. In particular, physical systems with the potential for parity-based topology are guaranteed to have Hamiltonians of size $2n$-by-$2n$. Thus, if such a system is 2D, any associated spectral localizer has size $4n$-by-$4n$ due to tensoring with the Pauli matrices. If the size of $H$ happens to be divisible by $4$ (not just $2$), the size of $L_{(\mathbf{x},E)}$ will be divisible by $8$, and the Pfaffian of $L_{(\mathbf{x},E)}$ in a skew-symmetric basis will be unaffected by multiplying by $\pm i$ as $(\pm i)^4 = 1$. However, even if the size of $L_{(\mathbf{x},E)}$ is only divisible by $4$, multiplying by $\pm i$ still yields an internally consistent sign of the Pfaffian, i.e., consistent across different choices of $(\mathbf{x},E)$. Thus, by choosing $(\mathbf{x},E)$ far outside of the system's spatial or spectral extent where the system is provably guaranteed to be trivial \cite{hastings_almost_2010}, the index value corresponding to a trivial material phase can be identified, and any changes in this index as $(\mathbf{x},E)$ is varied guarantee the appearance of nearby boundary-localized states. As such, rather than Eq.~\eqref{eq:local_index}, it is more efficient to compute
\begin{equation}
\xi^\textrm{L}_{(\mathbf{x},E)}(X, Y, H) = 
\textrm{sign} \left[ \textrm{Pf}\left( \pm i Q^\dagger L_{(\mathbf{x},E)}(X, Y, H) Q \right) \right] ,
\end{equation}
and, if necessary, perform a single additional calculation for large $\mathbf{x}$ or $E$, or some other location in position-energy space where the system is known to be trivial.


\section{$\mathbb{Z}_2$ topology in an aperiodic 2D class AII system}

\begin{figure}[t]
\center
\includegraphics[width=\columnwidth]{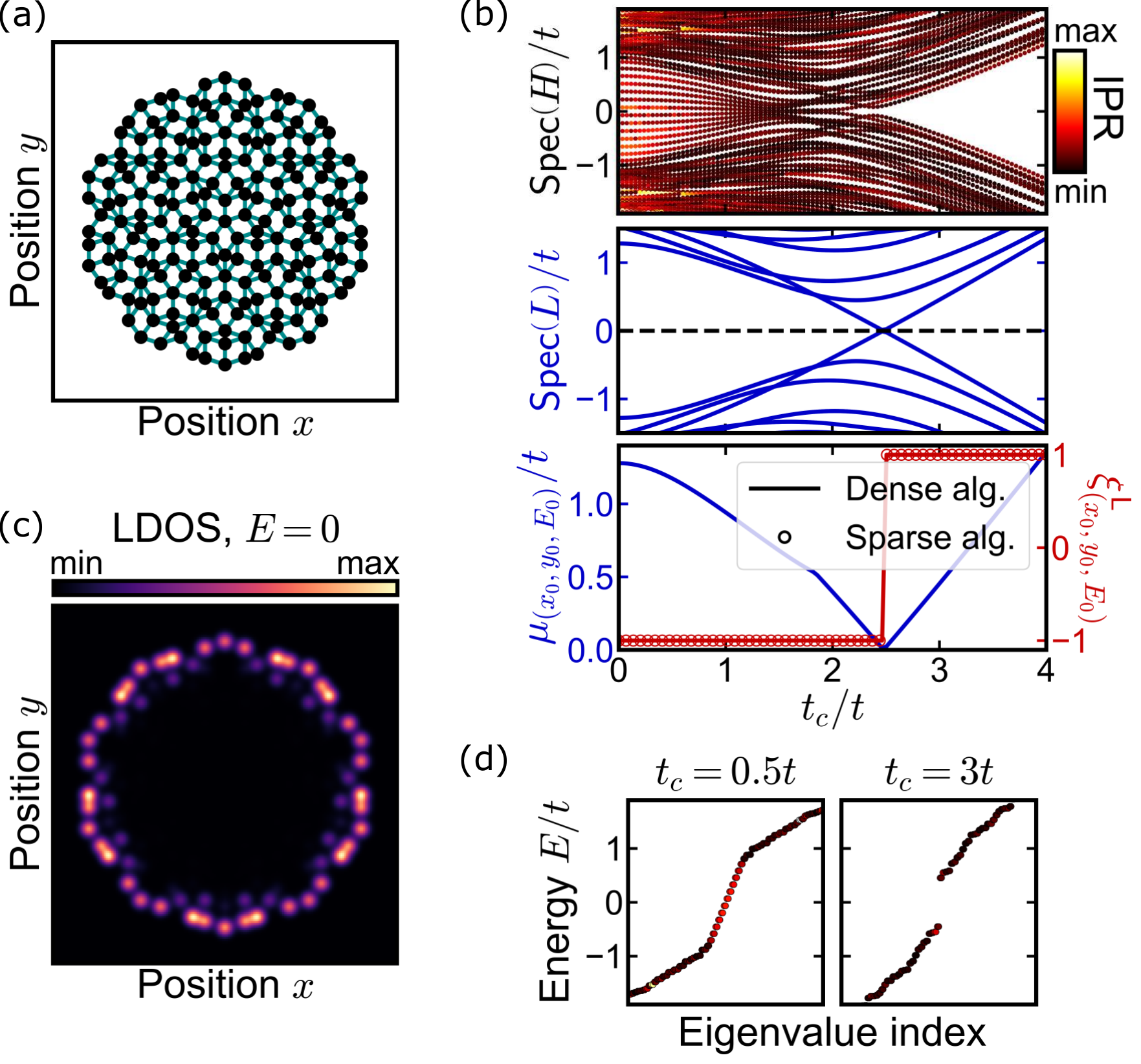}
\caption{
\textbf{Non-trivial $\mathbb{Z}_2$-topology in a quasicrystalline system.}
(a) Schematic of a Penrose tiling where the solid cyan lines represent the couplings between the internal degrees of freedom on the vertices of the tiling, with parameters $t=1$, $u = 2 t$, $t_1 = 2 t$, $\Delta = 2 t$, and $t_c = 0.3 t$.
(b) Spectrum of the system's Hamiltonian $\text{Spec} \left( H \right)$ and of the spectral localizer $\text{Spec} \left( L_{(x_0,y_0,E_0)} \right)$ against the inter-layer couplings $t_c$, and the associated local gap $\mu_{(x_0,y_0,E_0)}$ and local $\mathbb{Z}_2$-topological invariant $\xi^\text{L}_{(x_0,y_0,E_0)}$, calculated at the center of the lattice $(x_0,y_0)$ and energy $E_0=0$.
The colors in the $\text{Spec} \left( H \right)$ panel represent the inverse participation ratio (IPR).
(c) Local density of states (LDOS) at $E=0$. Each lattice site is represented as a two-dimensional Gaussian for better visualization.
(d) Spectrum of the system's Hamiltonian $\text{Spec} \left( H \right)$ for inter-layer coupling $t_c = 0.5 t$ and $t_c = 3 t$.
Spectral localizer-based calculations use $\kappa= 1.7 (t/a)$ with $a$ being the length between coupled sites.
}
\label{fig:qc_qshe}
\end{figure}

To demonstrate the application of the spectral localizer framework for classifying the topology in non-periodic system with fermionic time-reversal symmetry (class AII systems) and compare our sparse matrix factorization methods against existing dense matrix approaches, we consider a system composed of two coupled quasicrystalline layers that would independently possess opposite Chern numbers, similar to the Bernevig-Hughes-Zhang (BHZ)~\cite{Bernevig2006}.  
The single-layer Chern quasicrystals are constructed following a generalization of the Qi-Wu-Zhang (QWZ) model~\cite{Qi2006} to non-crystalline systems~\cite{Fulga2016} where the sites in the tight-binding model correspond to the vertices of a Penrose tiling and the couplings follow the edges of the tiles, as shown in Fig.~\ref{fig:qc_qshe}(a).
In particular, the Hamiltonian $H_1$ of the first quasicrystal layer has two internal degrees of freedom, where each vertex $j$ has a staggered on-site mass
\begin{equation}
\label{eq:H_qc_chern_jj}
[H_{1}]_{jj} = u \sigma_z
,
\end{equation}
and the couplings between the vertices $j$ and $k$ are given by
\begin{equation}
\label{eq:H_qc_chern_jk}
[H_{1}]_{jk} = t_1 \sigma_z + i \frac{1}{2} \Delta \left( \sigma_x \cos(\theta_{jk}) + \sigma_y \sin(\theta_{jk}) \right)
,
\end{equation}
with $\theta_{jk}$ being the angle of the bond between the vertices $j$ and $k$ with respect to the (positive) horizontal axis, and $t_1$ and $\Delta$ determine the couplings between the internal degree of freedom.
The Hamiltonian $H_2$ of the second layer is then given by $H_2 = H_1^*$.
Altogether, the two-layer system, which realizes a quantum spin Hall-like system without translational symmetry, is
\begin{equation}
\label{eq:H_qc_tr}
H = 
\left(
\begin{array}{cc}
H_1 & C \\
C^\dagger & H_1^*
\end{array}
\right)
,
\end{equation}
with $C = t_c I \otimes \sigma_y$ giving the inter-layer couplings acting on the internal degrees of freedom and chosen to satisfy fermionic time-reversal symmetry~\cite{Asboth2016}.
Note, the Hamiltonian in Eq.~\eqref{eq:H_qc_tr} has the same block form as in Eq.~\eqref{eq:H_tri} with $A = H_1$ and $B = C$.
%

%

For a range of couplings $t_c$ between the quasicrystalline layers, the aperiodic system possesses non-trivial $\mathbb{Z}_2$ topology at mid gap and at the lattice's center that can be classified using $\xi^\text{L}_{(x_0,y_0,0)}$, see Fig.~\ref{fig:qc_qshe}(b) (also see Supplementary Material Sect.~\ref{sect_supp:qc_chern} for further details~\cite{supp}).
This non-trivial $\mathbb{Z}_2$ phase also corresponds to the appearance of topological edge states at the boundary of the quasicrystalline system, as shown in the local density of states (LDOS) summed over all the degrees of freedom at $E=0$ in Fig.~\ref{fig:qc_qshe}(c) and in the in-gap states appearing in the finite system's full spectrum in Fig.~\ref{fig:qc_qshe}(d).
However, when the inter-layer coupling $t_c$ is too strong, the system becomes trivial with no topological edge states around $E=E_0$, see Fig.~\ref{fig:qc_qshe}(d).
As expected, the topological transition in the inter-layer coupling occurs at precisely the $t_c$ when the spectral gap of $L_{(\mathbf{x},E)}$ closes; 
as the coupling $t_c$ is increased, two eigenvalues simultaneously cross zero, forcing the local gap to close and resulting in a change of the local parity-based topological index.

\begin{figure}[t]
\center
\includegraphics[width=\columnwidth]{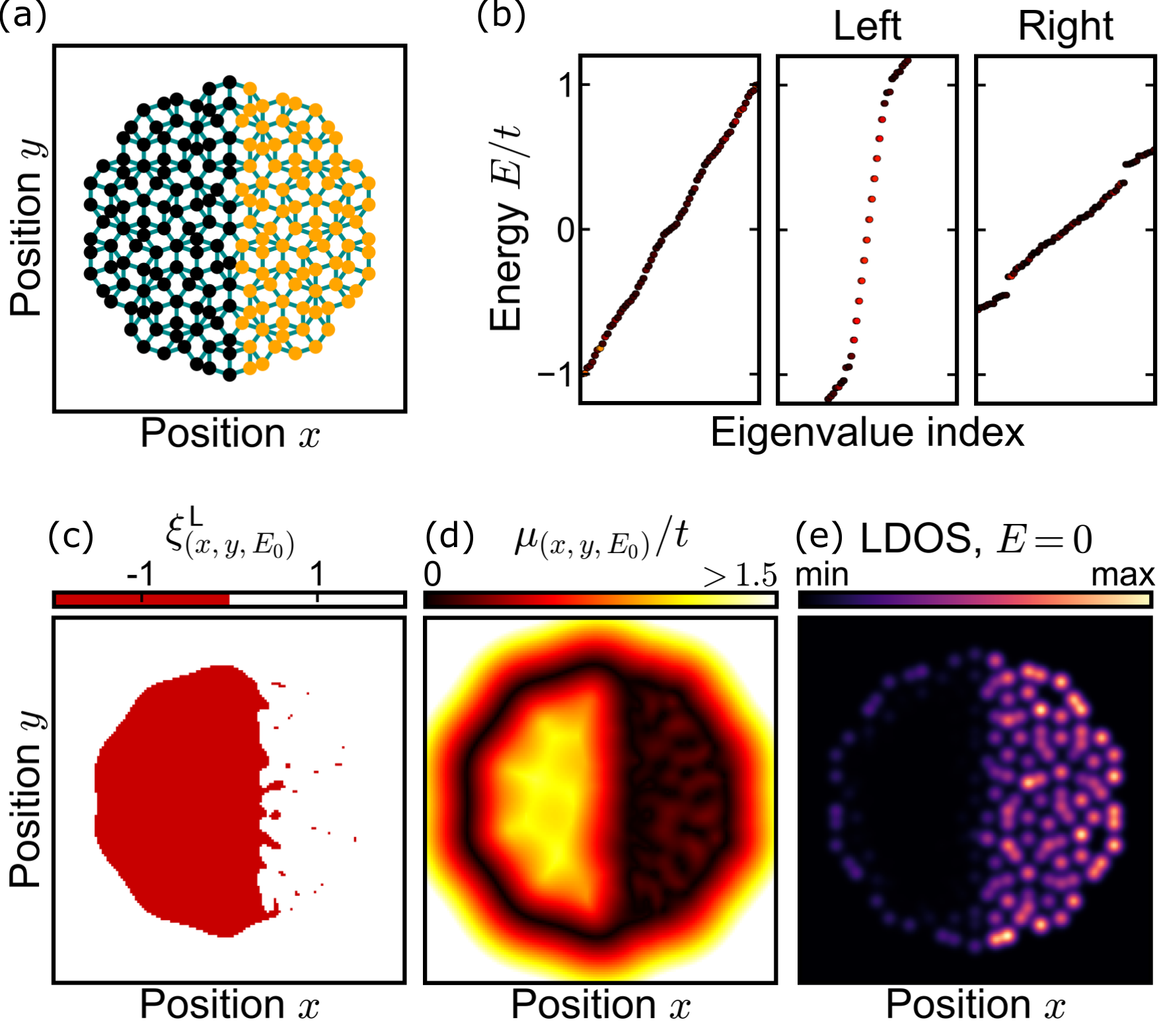}
\caption{
\textbf{$\mathbb{Z}_2$-topology in quasicrystalline gapless heterostructure system.}
(a) Penrose tiling where the solid cyan lines represent the couplings between the internal degree of freedom on the vertices of the tiling. 
The left part of the lattice with black dots has parameters $t=1$, $u = 2 t$, $t_1 = 1 t$, $\Delta = 2 t$ and $t_c = 0.3 t$, while the right part of the lattice with orange dots has parameters $t=1$, $u = 1 t$, $t_1 = 1 t$, $\Delta = 0.1 t$ and $t_c = 0$.
(b) Spectrum of the Hamiltonian of the heterostructure quasicrystal  lattice $\text{Spec} \left( H \right)$, as well as the spectra of Hamiltonians of the monostructure lattice using parameters of the left and right part in (a).
(c)-(d) Local invariant $\xi^\text{L}_{(x,y,E_0)}$ and local gap $\mu_{(x,y,E_0)}$ at energy $E_0=0$, respectively.
(e) Local density of states (LDOS) at $E=0$. Each lattice site is represented as a two-dimensional Gaussian for better visualization. 
Spectral localizer-based calculations use $\kappa= 1.7 (t/a)$ with $a$ being the length between coupled sites.
}
\label{fig:qc_hetero}
\end{figure}

Beyond the identification of the $\mathbb{Z}_2$ topology in gapped non-periodic systems, the spectral localizer framework also enables the investigation of topological properties in gapless systems.
To exemplify the topological classification in an aperiodic gapless system, we consider a gapless quasicrystalline heterostructure as shown in Fig.~\ref{fig:qc_hetero}.
In particular, the system's left side uses the same parameters as in Fig.~\ref{fig:qc_qshe} with $t_c = 0.3 t$, and is thus gapped with non-trivial $\mathbb{Z}_2$ topology, while the system's right side is chosen to be gapless, as revealed by the corresponding spectra plotted in Fig.~\ref{fig:qc_hetero}(b).
Altogether, the gapless spectrum of the whole quasicrystalline heterostructure system is shown in the left panel of Fig.~\ref{fig:qc_hetero}(b).    
Despite the system's lack of a bulk spectral gap common to both domains, both the local $\mathbb{Z}_2$ index $\xi^\text{L}_{(x,y,E_0)}$ and the local gap $\mu_{(x,y,E_0)}$ over the whole system at $E_0=0$ can be computed from the spectral localizer, as shown in Fig.~\ref{fig:qc_hetero}(c).
Furthermore, the variation of the local index as a function of position is inevitably associated to the closing of the local gap $\mu_{(x,y,E_0)} \rightarrow 0$, as seen in Fig.~\ref{fig:qc_hetero}(d).
As such, while the system's LDOS at $E=0$ [Fig.~\ref{fig:qc_hetero}(c)] is not conclusive in confirming the topological nature of the boundary-localized state, the local gap's closing guarantees the existence of localized boundary states near the locations where $\mu_{(x,y,E_0)} \rightarrow 0$. 
Overall, the absence of a band gap results in ``leaky'' topological boundary modes that can be nevertheless identified and predicted within the spectral localizer framework.
Altogether, as apposed to projector-based approaches to identifying local topology, the spectral localizer is able to diagnose the topology over the full parameter space, even when the spectral gap of the Hamiltonian, a projected spin operator, or any other relevant operator vanish (see Supplementary Material Sect.~\ref{sect_supp:limit_proj} for further details~\cite{supp}).
%


\section{Fragile topology in a photonic quasicrystal}

\begin{figure*}[t]
\center
\includegraphics[width=2\columnwidth]{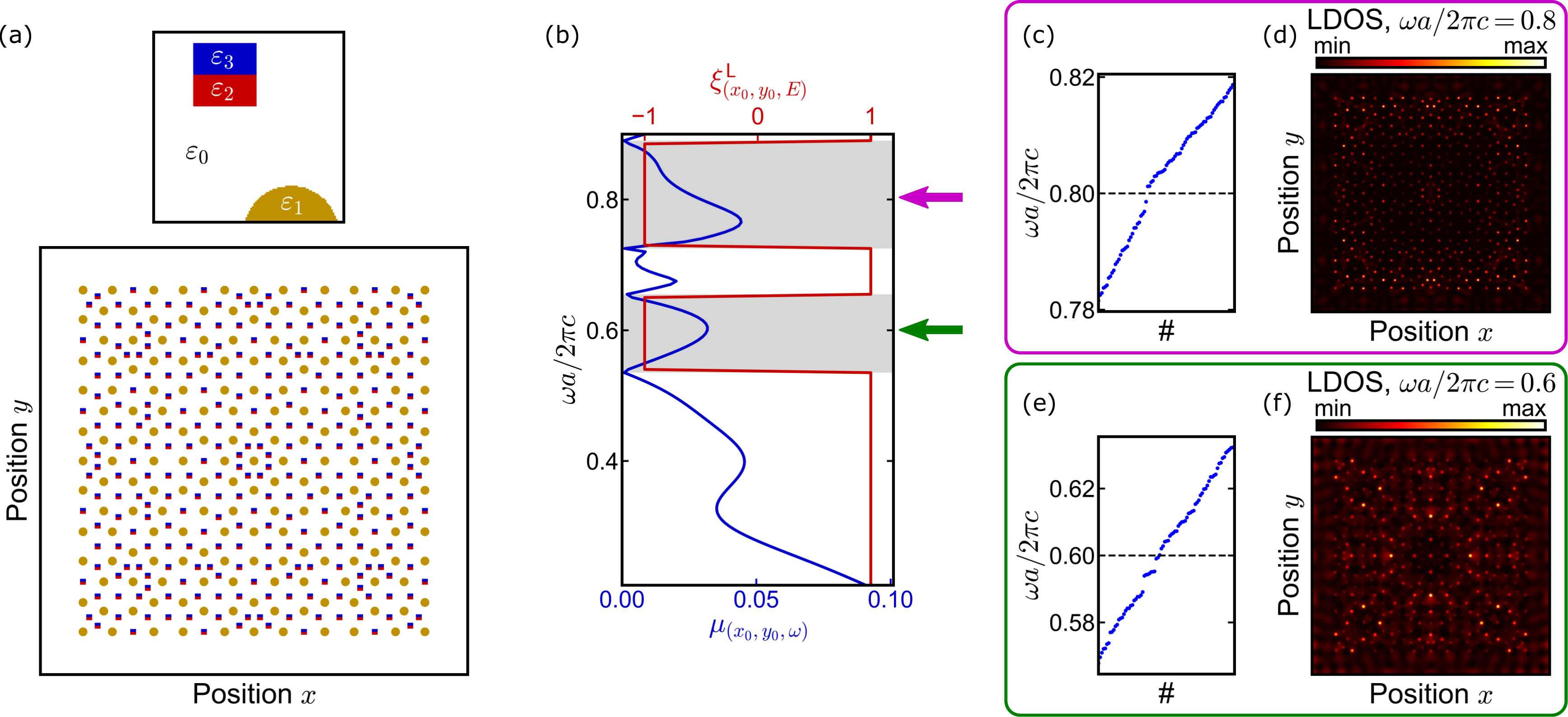}
\caption{
\textbf{Fragile topology in photonic quasicrystal.}
(a) 2D photonic quasicrystal constructed from the Ammann-Beenker quasicrystal. The dielectric rods are located on the vertices of the Ammann-Beenker quasicrystal, and the dielectric squares are located on the vertices of the Voronoi edges of the Ammann-Beenker quasicrystal. 
The zoom-in shows the different dielectric where $\varepsilon_0=1$, $\varepsilon_1=11.56$, $\varepsilon_{2,ii}=16$, $\varepsilon_{2,xy}=-\varepsilon_{2,yx}=6i$, $\varepsilon_3=\varepsilon_2^*$. Diameter of the rods is $r=0.15 a$ and edge length of the squares is $d=0.2 a$, with $a$ being the side length of the square tile in the Ammann-Beenker quaaicrystal. 
(b) Local gap $\mu_{(x_0,y_0,\omega)}$ and local $\mathbb{Z}_2$-topological invariant $\xi^\text{L}_{(x_0,y_0,\omega)}$ as the normalized frequency $\omega a/2\pi c$ is varied, calculated at the center of the lattice $(x_0,y_0)$. 
The gray shaded areas show the fragile topological region with non-trivial $\mathbb{Z}_2$ index.
(c) Spectrum of the photonic system around $\omega a/2\pi c = 0.8$.
(d) Local density of states (LDOS) at $\omega a/2\pi c = 0.8$.
(e)-(f) Same as (c)-(d) but for $\omega a/2\pi c = 0.6$.
Spectral localizer-based calculations use $\kappa = 0.04 (2\pi c/a^2)$. 
}
\label{fig:qc_phc_fragile}
\end{figure*}

To demonstrate the full advantage of a sparse implementation to finding a skew-symmetric matrix's Pfaffian's sign, we turn to the classification of fragile topology in a 2D photonic quasicrystal that is $C_2 \mathcal{T}$-symmetric, where now $\mathcal{T}^2 = I$ refers to bosonic time-reversal symmetry. In particular, we here consider a quasicrystal where the dielectric rods of permittivity $\varepsilon_1$ are located at the vertices of a Ammann-Beenker tiling, and dielectric squares are placed on the vertices of the Voronoi edges of the tiling, see Fig.~\ref{fig:qc_phc_fragile}(a). Furthermore, the dielectric squares are divided into two different materials that both exhibit the gyro-optical effect, such that $\varepsilon_2$, and $\varepsilon_3 = \varepsilon_2^*$. This design is inspired from the periodic square photonic lattice of rods in Ref.~\cite{Lee2025}, where the dielectric squares are on the vertices of the Voronoi edges of the square lattice. Altogether, this 2D photonic Ammann-Beenker quasicrystal is chosen to be $C_2 \mathcal{T}$-symmetric, but while having individually broken $C_2$ and $\mathcal{T}$ symmetries.

To classify the topology of the gyro-optic 2D photonic quasicrystal, we use established methods to first convert Maxwell's equations into a Hamiltonian, use a finite-difference discretization to convert the partial differential equations into a matrix, and then use this matrix in the spectral localizer framework~\cite{Lee2025,Cerjan2022a}. For the transverse electric (TE) sector of a 2D photonic structure with nonzero $[H_z(\mathbf{x}),E_x(\mathbf{x}),E_y(\mathbf{x})]^\top$, Maxwell's equations for time-independent linear materials can be written as a generalized eigenvalue equation for the frequency $\omega$ as 
\begin{multline}
    \left[\begin{array}{ccc}
    0 & i \partial_y & -i \partial_x \\
    i \partial_y & 0 & 0 \\
    -i \partial_x & 0 & 0
    \end{array}
    \right] \left[\begin{array}{c}
    H_z(\mathbf{x}) \\
    E_x(\mathbf{x}) \\
    E_y(\mathbf{x})
    \end{array}
    \right] \\
    = \omega
    \left[\begin{array}{ccc}
    \mu_{zz}(\mathbf{x}) & 0 & 0 \\
    0 & \varepsilon_{xx}(\mathbf{x}) & \varepsilon_{xy}(\mathbf{x}) \\
    0 & \varepsilon_{yx}(\mathbf{x}) & \varepsilon_{yy}(\mathbf{x})
    \end{array}
    \right] \left[\begin{array}{c}
    H_z(\mathbf{x}) \\
    E_x(\mathbf{x}) \\
    E_y(\mathbf{x})
    \end{array}
    \right],
\end{multline}
with $\varepsilon_{yx}(\mathbf{x}) = \varepsilon_{xy}^*(\mathbf{x})$.
Here, we assume that the permittivity $\varepsilon$ and permeability $\mu$ are both sufficiently constant and loss-less over a frequency range of interest, and that we are considering dielectrics for which $\varepsilon$ and $\mu$ are both positive. For such cases, the effective Hamiltonian can be written as
\begin{multline}
    H(\mathbf{x}) = \left[\begin{array}{ccc}
    \mu_{zz}(\mathbf{x}) & 0 & 0 \\
    0 & \varepsilon_{xx}(\mathbf{x}) & \varepsilon_{xy}(\mathbf{x}) \\
    0 & \varepsilon_{yx}(\mathbf{x}) & \varepsilon_{yy}(\mathbf{x})
    \end{array}
    \right]^{-\frac{1}{2}} \\
    \times \left[\begin{array}{ccc}
    0 & i \partial_y & -i \partial_x \\
    i \partial_y & 0 & 0 \\
    -i \partial_x & 0 & 0
    \end{array}
    \right] \left[\begin{array}{ccc}
    \mu_{zz}(\mathbf{x}) & 0 & 0 \\
    0 & \varepsilon_{xx}(\mathbf{x}) & \varepsilon_{xy}(\mathbf{x}) \\
    0 & \varepsilon_{yx}(\mathbf{x}) & \varepsilon_{yy}(\mathbf{x})
    \end{array}
    \right]^{-\frac{1}{2}}
\end{multline}
where $M^{-1/2}$ denotes the principal square root of a positive $M$. A global $H$ for a full finite system can then be recovered by choosing a finite-difference discretization~\cite{yee_numerical_1966,taflove_advances_2013} or a finite-element mesh~\cite{Wong2024}. Here, we make use of a finite-difference approach with perfect electric conductor (PEC) boundary conditions, which are Dirichlet boundary conditions on the $\mathbf{E}$ field.
Finally, the spectral localizer framework can classify fragile topology stemming from $C_2 \mathcal{T}$ symmetry using
\begin{align}
\begin{split}
L&_{(x,y,\omega)}(WXW^\dagger,WYW^\dagger,WHW^\dagger) \\
&= \kappa(WXW^\dagger-xI)\otimes\sigma_x+ \kappa(WYW^\dagger-yI)\otimes\sigma_z \\
&\quad + (WHW^\dagger-\omega I)\otimes\sigma_y,  \label{eq:fragileL}
\end{split}
\end{align}
such that the parity-based energy-resolved invariant is
\begin{multline}
\zeta_E(X,Y,H) \\
= \textrm{sign}[\textrm{Pf}(iL_{(0,0,\omega)}(WXW^\dagger,WYW^\dagger,WHW^\dagger))], \label{eq:fragileI}
\end{multline}
where the $i$ has been inserted following the discussion at the end of Sec.~\ref{sec:3b}.
Here, $W = (1/\sqrt{2})\left(C_2+iI\right)$ is the unitary matrix that transforms $H$ to be skew-symmetric while preserving $X$ and $Y$ to be symmetric and the invariant must be evaluated at the finite system's rotational center $\mathbf{x} = (0,0)$, so that $L_{(x,y,\omega)}^\dagger = L_{(x,y,\omega)}$ and $L_{(0,0,\omega)}^\top = -L_{(0,0,\omega)}$

Using Eqs.~\eqref{eq:fragileL} and \eqref{eq:fragileI}, alongside Eq.~\eqref{eq:local_gap}, we can classify the fragile topology of a large 2D photonic quasicrystal based on the Ammann-Beenker tiling, see Fig.~\ref{fig:qc_phc_fragile}. Due to the size of the system and the need to choose a sufficiently dense discretization, the resulting $L_{(0,0,\omega)}$ has size $\approx (3.2 \cdot 10^6)$-by-$(3.2 \cdot 10^6)$. Note that $L_{(0,0,\omega)}$ is extremely sparse, but it is not banded; thus, it is not possible to calculate the Pfaffian of $L_{(0,0,\omega)}$ directly using existing approaches for either dense or sparse-banded matrices~\cite{wimmer2011efficient,grabsch2019pfaffian}. Nevertheless, our factorization approach is able to tackle each classification problem of a 3 million unknown problem at a different $\omega$ on a standard workstation in 1-2 hours. Moreover, the topological phase transitions identified by the energy-resolve fragile index coincide with vanishing local gaps $\mu_{(0,0,\omega)}$, which are independently calculated using a standard sparse eigenvalue solver, confirming the accuracy of our pivot and factorization strategy.


\section{Discussion}

In this work, we have demonstrated how the spectral localizer framework enables efficient classification of parity-based $\mathbb{Z}_2$ topology in aperiodic systems, addressing several long-standing limitations of existing topological classification approaches. By exploiting the intrinsic skew-symmetric structure of the spectral localizer due to the system's underlying physical symmetries, we showed that topological information can be extracted from the sign of a Pfaffian without the need to use spectral projectors, band gaps, or spin-resolved subspaces. From a physical perspective, this establishes a robust, energy- and position-resolved notion of topology that remains meaningful even in gapless or inhomogeneous systems, while from a mathematical standpoint it connects real-space topological classification directly to homotopy classes of large sparse operators. The accompanying local gap provided by the spectral localizer plays a dual role, it both quantifies topological protection and enforces a bulk–boundary correspondence by guaranteeing the appearance of approximately localized states where the invariant changes value.

A key technical contribution of this work is the development of a scalable numerical algorithm for determining the sign of the Pfaffian of extremely large sparse skew-symmetric matrices, which enables the practical application of these ideas to realistic material models far beyond the reach of dense linear algebra methods. Our examples of classifying parity-based topology in quasicrystalline quantum spin Hall systems and fragile photonic quasicrystals demonstrate the broad range of utility of the spectral localizer framework and our sparse sign Pfaffian algorithm. As such, we anticipate that these methods will be particularly valuable for the systematic discovery of topological phases in complex material platforms—such as twisted van der Waals systems and large-scale photonic or acoustic metamaterials—where translational symmetry is absent and conventional momentum-space diagnostics fail.

All the $\mathbb{Z}$ and $\mathbb{Z}_2$ indices derived from the spectral localizer rely on a sparse matrix factorization, either LU, LDLT, or the skew-LDLT discussed here.  While these algorithms are not guaranteed to give a sparse output \cite{duff_number_1974,duff_note_1983}, in practice our calculations show that they give reasonably sparse results when applied to models of physically realizable topological systems.  For such realistic models of systems in 2D and 3D that are large enough to avoid finite-size effects, one is  constrained to only use sparse matrix methods.  Thus, it is anticipated that the sparse factorization methods implemented here will have applications to the prediction of material properties more broadly, beyond the computation of the local index.


\section*{Acknowledgments}

\textbf{Data and Code Availability:}
The code that this study is based on is publicly available in a GitHub repository \cite{wong_sparse_sign_pfaffian_2026}. This repository also contains the data files needed to reproduce the results in Fig.~\ref{fig:qc_qshe} and \ref{fig:qc_phc_fragile}.

\textbf{Funding:} 
S.W., I.Y., and C.S.\ acknowledge support from the Laboratory Directed Research and Development program at Sandia National Laboratories.
Research by T.L.~was sponsored by the Army Research Office and was accomplished under Grant Number W911NF-25-1-0052. 
A.C.\ acknowledges support from the U.S.\ Department of Energy, Office of Basic Energy Sciences, Division of Materials Sciences and Engineering.
This work was performed in part at the Center for Integrated Nanotechnologies, an Office of Science User Facility operated for the U.S. Department of Energy (DOE) Office of Science.
Sandia National Laboratories is a multimission laboratory managed and operated by National Technology \& Engineering Solutions of Sandia, LLC, a wholly owned subsidiary of Honeywell International, Inc., for the U.S. DOE's National Nuclear Security Administration under Contract No. DE-NA-0003525. 
%
The views and conclusions contained in this document are those of the authors and should not be interpreted as representing the official policies, either expressed or implied, of the Army Research Office or the U.S.~Government. The U.S.~Government is authorized to reproduce and distribute reprints for Government purposes notwithstanding any copyright notation herein. The views expressed in the article do not necessarily represent the views of the U.S.~DOE or the United States Government.

\bibliography{ref}

\appendix
\renewcommand{\appendixname}{Supplemental Material}
\renewcommand{\thesection}{\Roman{section}}

\section{Pfaffians and homotopy of skew-symetric matrices}
\label{sect_supp:Pfaffians}

The matrices of a given size that are Hermitian, skew-adjoint and invertible form a set of two connected components corresponding to fermionic parity.  Kitaev \cite{kitaev2001unpaired_Majorana} defined this as the sign of the Pfaffian.  From a mathematical point of view it is simpler to multiply these matrices by $i$ and so we consider instead the set of invertible skew-adjoint real matrices of size $2n$-by-$2n$.  We derive what we need about Pfaffians along the way.

One component will consists of matrices that can be connected to 
\begin{equation}
J_{+}=\left[\begin{array}{cccc}
J_{0}\\
 & J_{0}\\
 &  & \ddots\\
 &  &  & J_{0}
\end{array}\right],\label{eq:Canonical_pf_positive}
\end{equation}
where $J_0 = i\sigma_y$, 
and the second will consist of those that can be connected to 
\begin{equation}
J_{-}=\left[\begin{array}{cccc}
-J_0\\
 & J_0\\
 &  & \ddots\\
 &  &  & J_0
\end{array}\right].\label{eq:Canonical_pf_negative}
\end{equation}

For notation, the space we are interested we will denote as
\begin{equation}
\textup{Sk}(2n,\mathbb{R})=\left\{ \left.X\in\mathbf{M}_{2n}(\mathbb{R})\right|X^{\top}=-X\right\} .
\end{equation}
We are especially interested in the invertible elements here, so define
\begin{equation}
\textup{Sk}^{-1}(2n,\mathbb{R})=\left\{ \left.X\in\textup{Sk}(2n,\mathbb{R})\right|X^{-1}\text{ exists}\right\} .
\end{equation}

\begin{lem}
\label{lem:spec_thm}
If $X$ in $\textup{Sk}(2n,\mathbb{R})$ then there is an orthogonal
matrix $U$ with
\begin{equation}
X=UTU^{\top}\label{eq:spectral_theorem}
\end{equation}
where $T$ is block-diagonal, with blocks
\begin{equation*}
\left[\begin{array}{cc}
0 & b_{j}\\
-b_{j} & 0
\end{array}\right].
\end{equation*}
Moreover the spectrum of $X$ is
$\left\{ \pm ib_{1},\cdots,\pm ib_{n}\right\}$ .
\end{lem}
\vspace{5px}
\noindent 
\textbf{Proof}.
Since $X$ is normal and real, this follows from the real-version
of the spectral theorem for normal matrices (Theorem 2.5.8 in \cite{Horn_Jonhson2013Matrix_Analysis}, for example). For a more explicit proof, see the
appendix to \cite{Cerjan2024a}. 

\vspace{5px}

\begin{thm}
\label{thm:Path_to_a_standard_matrix}If $X$ is in $\textup{Sk}^{-1}(2n,\mathbb{R})$
then there is a path in $\textup{Sk}^{-1}(2n,\mathbb{R})$ from $X$
to either $J_{+}$ or $J_{-}$.
\end{thm}

\vspace{5px}
\noindent 
\textbf{Proof}
We can modify the factorization in Equation~\ref{eq:spectral_theorem}
by negating one column of $U$ and one row and one column of $X$
to flip the sign of $\det(U)$. Thus we can assume $\det(U)=1$. Any orthogonal matrix of determinant one can be
connected to $I$ by a path $U_{t}$ of unitary matrices. Thus $U_{t}TU_{t}^{\top}$
is a path of skew-symmetric real matrices from $X$ to $T$. We can
move the $\lambda_{j}$ individually through nonzero real numbers
until $\lambda_{j}=\pm1$.

The skew-symmetric real matrices
\begin{equation*}
W_{1}=\left[\begin{array}{cccc}
0 & \mp1\\
\pm1 & 0\\
 &  & 0 & -1\\
 &  & 1 & 0
\end{array}\right],\ W_{2}=\left[\begin{array}{cccc}
0 & \pm1\\
\mp1 & 0\\
 &  & 0 & 1\\
 &  & -1 & 0
\end{array}\right]
\end{equation*}
can be connected through invertible real skew-symmetric matrices, specifically
\begin{equation*}
\left[\begin{array}{cccc}
0 & \mp\cos\theta &  & \sin\theta\\
\pm\cos\theta & 0 & \sin\theta\\
 & -\sin\theta & 0 & -\cos\theta\\
-\sin\theta &  & \cos\theta & 0
\end{array}\right].
\end{equation*}
Padding this with ones on the diagonal we can find a homotopy from
$T$ to a new matrix with two signs flipped on the super-diagonal.
We can do this until the super diagonal is all ones except perhaps
a $-1$ in the top-most position.

\vspace{5px}

To finish, we need to show that the matrices $J_{+}$ and $J_{-}$  cannot be connected by a path that stays invertible and skew-symmetric.
One way to do this is to find a continuous function from this space of matrices onto $\mathbb{R}\setminus\{0\}$. As the punctured real line has two connected components, the space $\textup{Sk}^{-1}(2n,\mathbb{R})$
must have at least two components as well. The function 
\begin{equation*}
\det:\textup{Sk}^{-1}(2n,\mathbb{R})\rightarrow\mathbb{R}\setminus\{0\}
\end{equation*}
is promising, as it is continuous (indeed a polynomial) in the entrees of the input matrix. This approach fails, as Lemma~\ref{lem:spec_thm} tells us that the product of all eigenvalues, which equals the determinant, is always positive for a real, skew-symmetric matrix.

What we want is a continuous square root of the determinant. One of these is the Pfaffian. We find it using factorization of polynomials, following similar approaches in \cite{Lax2007LinearAlgebra} and \cite{Serre2010Matrices}.

The fact that the Pfaffian is a polynomial in the entries in the matrix tells us that the Pfaffian is probably no harder to compute than a determinant.  When using the Pfaffian, what is important it that it varies differentiably with respect to the input.

\begin{lem}
\label{lem:Det_of_tridiagonal}
If $T$ is a tridiagonal skew-symmetric matrix in
$\mathbf{M}_{2n}(\mathbb{C})$,
\begin{equation*}
T=\left[\begin{array}{cccccc}
0 & t_{1}\\
-t_{1} & 0 & t_{2}\\
 & -t_{2} & 0 & t_{3}\\
 &  & -t_{3} & 0 & \ddots\\
 &  &  & \ddots & \ddots & t_{2n-1}\\
 &  &  &  & -t_{2n-1} & 0
\end{array}\right]
\end{equation*}
then
\begin{equation*}
\det(T)=\Biggl(\prod_{j=1}^{n}t_{2j-1}\Biggr)^{2}.
\end{equation*}
\end{lem}

\vspace{5px}
\noindent 
\textbf{Proof}
We expand on the first row, then the first column:
\begin{equation*}
\det(T)=\left(-t_{1}\right)\left(-t_{1}\right)\det\left[\begin{array}{cccc}
0 & t_{3}\\
-t_{3} & 0 & \ddots\\
 & \ddots & \ddots & t_{2n-1}\\
 &  & -t_{2n-1} & 0
\end{array}\right].
\end{equation*}
The result now follows, by induction.

\vspace{5px}

\begin{lem}
\label{lem:poly_column_fixer} Suppose $A$ is a skew-symmetric matrix in $\mathbf{M}_{n}(\mathbb{C})$. There is a matrix $C=C(A)$ whose elements are real polynomials in the elements of $A$ such that $CAC^{\top}$ has zeros in the first column except in the second row and such that $\det(C(A))$ is not the zero
polynomial.
\end{lem}

\vspace{5px}
\noindent 
\textbf{Proof}
Let's label the first row and column of $A$, so
\begin{equation*}
A=\left[\begin{array}{cccccc}
0 & a_{2} & a_{3} & a_{4} & \cdots & a_{n}\\
-a_{2} & 0 & * & * & \cdots & *\\
-a_{3} & * & 0 & * & \cdots & *\\
-a_{4} & * & * & 0\\
\vdots & \vdots & \vdots &  & \ddots\\
-a_{n} & * & * &  &  & *
\end{array}\right].
\end{equation*}
Let 
\begin{equation*}
C(A)=C=\left[\begin{array}{cccccc}
1\\
0 & 1\\
 & -a_{3} & a_{2}\\
 &  & -a_{4} & a_{3}\\
 &  &  & \ddots & \ddots\\
 &  &  &  & -a_{n} & a_{n-1}
\end{array}\right].
\end{equation*}
We see that $CA$ will have the same first-two rows as $A$ but the
first column will be
\begin{equation*}
\left[\begin{array}{cccccc}
1\\
0 & 1\\
 & -a_{3} & a_{2}\\
 &  & -a_{4} & a_{3}\\
 &  &  & \ddots & \ddots\\
 &  &  &  & -a_{n} & a_{n-1}
\end{array}\right]\left[\begin{array}{c}
0\\
-a_{2}\\
-a_{3}\\
-a_{4}\\
\vdots\\
-a_{n}
\end{array}\right]=\left[\begin{array}{c}
0\\
-a_{2}\\
0\\
0\\
\vdots\\
0
\end{array}\right].
\end{equation*}
By symmetry, we see that 
\begin{equation*}
CAC^{\top}=\left[\begin{array}{cccccc}
0 & a_{2} & 0 & 0 & \cdots & 0\\
-a_{2} & 0 & * & * & \cdots & *\\
0 & * & 0 & * & \cdots & *\\
0 & * & * & 0\\
\vdots & \vdots & \vdots &  & \ddots\\
0 & * & * &  &  & *
\end{array}\right].
\end{equation*}
Since $C$ is lower-triangular, $\det(C)$ will equal
$a_{2}a_{3}\cdots a_{n}$.

\vspace{5px}

\begin{thm}
If $X$ is a skew-symmetric real matrix, $2n$-by-$2n$, then $\det(X)$
equals the square of a real  polynomial in the
elements of $X$. 
\end{thm}

\vspace{5px}
\noindent 
\textbf{Proof}
Let $m=2n$. If we apply Lemma~\ref{lem:poly_column_fixer} $m-2$
times we find $C_{j}$ whose elements are real
polynomials in the elements of $X$ such that $\det(C_{j})$ is not
the zero polynomial and
\begin{equation}
C_{m-2}\cdots C_{2}C_{1}XC_{1}^{\top}C_{2}^{\top}\cdots C_{m-2}^{\top}\label{eq:CXC}
\end{equation}
is tridiagonal. Lemma~\ref{lem:Det_of_tridiagonal} tells us that the determinant here is the square of a real polynomial, so we have
\begin{equation}
\det(X)\left(p(x_{11},\dots,x_{mm})\right)^{2}=\left(q(x_{11},\dots,x_{mm})\right)^{2}\label{eq:Det(C)Det(X)Det(C)_formula}
\end{equation}
where $p$ equals the product of the determinants of the $C_{j}$
and $q^{2}$ is the determinant of the matrix in Eq.~\ref{eq:CXC}. The ring $\mathbb{R}[x_{11},\dots,x_{mm}]$ of all polynomials in
$m^{2}$ commuting indeterminants with real coefficients is a unique factorization domain (see Ch.~18 of \cite{judson2022abstract}). Thus the polynomial $\det(X)$ must factor into a product of squares of irreducible
polynomials. It is therefore the square of a polynomial. 

\vspace{5px}

\begin{cor}
The matrices $J_{+}$ and $J_{-}$ cannot be connected by a path in
$\textup{Sk}^{-1}(2n,\mathbb{R})$.
\end{cor}

\vspace{5px}
\noindent 
\textbf{Proof}
Let $p(x_{11},\dots,x_{mm})$ denote a polynomial in the entries of
$X$ such that 
\begin{equation}
\left(p(x_{11},\dots,x_{mm})\right)^{2}=\det(X).\label{eq:p^2_is_det}
\end{equation}
We will use the notation $p(X)$ to mean $p(x_{11},\dots,x_{mm})$.
Consider first the obvious path
\begin{equation*}
X(t)=\left[\begin{array}{ccccc}
tJ\\
 & J\\
 &  & J\\
 &  &  & \ddots\\
 &  &  &  & J
\end{array}\right]
\end{equation*}
so that $X(1)=J_{+}$ and $X(-1)=J_{-}$. Of course, this path becomes
singular at $t=0$, but all the elements in $X(t)$ are polynomials
in $t$. We know
\begin{equation*}
\det(X(t))=t^{2}.
\end{equation*}
Since $p(X(t))$ is a polynomial in $t$ we see that $p(X(t))=t$
or $p(X(t))=-t$. Thus $p(X(0))$ and $p(X(1))$ have opposite signs. 

Now suppose $Y(t)\in\textup{Sk}^{-1}(2n,\mathbb{R})$ is a path from
$X(0)$ to $X(1)$. Since $p(Y(0))$ and $p(Y(1))$ are real numbers
of opposite signs, by continuity there is some $t_{0}$ where $p(Y(t_{0}))=0$.
This means that $\det(Y(t_{0}))=0$ and so $Y(t_{0})$ is singular.
This contradiction proves that no such path exists.

\vspace{5px}

We have established the following.

\begin{thm}
For all $n$, the set $\textup{Sk}^{-1}(2n,\mathbb{R})$ of invertible,
real, skew-adjoint $2n$-by-$2n$ matrices has two connected components:
 those matrices path connected in $\textup{Sk}^{-1}(2n,\mathbb{R})$
to $J_{+}$, and those path connected to $J_{-}$.
\end{thm}

Let us denote these two connected components as $\textup{Sk}_{+}^{-1}(2n,\mathbb{R})$
and $\textup{Sk}_{-}^{-1}(2n,\mathbb{R})$. There are exactly four
choices for continuous functions $p$ from $\textup{Sk}(2n,\mathbb{R})$ to $\mathbb{R}$
that satisfy
$\left(p(X)\right)^{2}=\det(X)$.
They all must send singular matrices to zero. We can select a sign
for 
\begin{equation*}
p(X)=\pm\sqrt{\det(X)},\quad X\in\textup{Sk}_{+}^{-1}(2n,\mathbb{R})
\end{equation*}
and independently select a sign for 
\begin{equation*}
p(X)=\pm\sqrt{\det(X)},\quad X\in\textup{Sk}_{-}^{-1}(2n,\mathbb{R}).
\end{equation*}
If we select the same sign for both the function $p$ will not be
polynomial. The other two choices are polynomial. We break the tie
by selecting $+$ for matrices in $\textup{Sk}_{+}^{-1}(2n,\mathbb{R})$
and that polynomial function is known as the Pfaffian.

\vspace{5px}
\noindent 
\textbf{Def.}
If $X$ is in $\textup{Sk}(2n,\mathbb{R})$ then the \emph{Pfaffian}
of $X$ is defined as follows. If $X$ is in $\textup{Sk}_{\pm}^{-1}(2n,\mathbb{R})$
then 
\begin{equation*}
\textup{Pf }(X)=\pm\sqrt{\det(X)}
\end{equation*}
and $\textup{Pf }(X)=0$ if $X$ is singular.
\vspace{5px}

We prove the following, above. 

\begin{thm}
If $p(X)$ is a polynomial function, with integer coefficients, defined on $\textup{Sk}(2n,\mathbb{R})$
that satisfies
\[
\left(p(X)\right)^{2}=\det(X)
\]
for all $X$ in $\textup{Sk}(2n,\mathbb{R})$, then either $p(X)=\textup{Pf}(X)$,
for all $X$ in $\textup{Sk}(2n,\mathbb{R})$ or $p(X)=-\textup{Pf}(X)$
for all $X$ in $\textup{Sk}(2n,\mathbb{R})$.
\end{thm}

Now we derive several established formulas for computing Pfaffians efficiently via matrix factorization. 

\begin{lem}
\label{lem:rwo-column-swap-formula} If $U$ is the real orthogonal
matrix that swaps the first two elements in the standard basis of
$\mathbb{C}^{2n}$, and leaves all other elements in this basis fixed,
then
\begin{equation*}
\textup{Pf}(UXU^{\top})=-\textup{Pf}(X).
\end{equation*}
\end{lem}

\vspace{5px}
\noindent 
\textbf{Proof}
Since replacing $X$ by $UXU^{\top}$ only moves the elements of $X$
around, the function 
\begin{equation*}
p(X)=\textup{Pf}(UXU^{\top})
\end{equation*}
is again a polynomial. Since
\begin{equation*}
\left(p(X)\right)^{2}=\det(UXU^{\top})=\det(X)
\end{equation*}
we know $p(X)=\pm\textup{Pf}(X)$. Since $UJ_{+}U^{\top}=J_{-}$
we must have $p(X)=-\textup{Pf}(X)$.

\vspace{5px}

\begin{thm}
\label{thm:Det_Pfaff_of_AXA^T} If $A$ and $X$ are real $2n$-by-$2n$
matrices and $X$ is skew-symmetric, then
\begin{equation*}
\textup{Pf}\left(AXA^{\top}\right)=\det(A)\textup{Pf}(X).
\end{equation*}
\end{thm}

\vspace{5px}
\noindent 
\textbf{Proof}
If $\det(A)=0$ then $A$ and $AXA^{\top}$ are singular and the formula
is true as it only says $0=0$.

Next, let's assume that $\det(A)$ is positive. Consider the function
on $\textup{Sk}(2n,\mathbb{R})$ defined by 
\begin{equation*}
p(X)=\textup{Pf}\left(AXA^{\top}\right)/\det(A).
\end{equation*}
This is some polynomial in $x_{11},\dots,x_{mm}$ and we compute
\begin{align*}
\left(p(X)\right)^{2} & =\left(\textup{Pf}\left(AXA^{\top}\right)\right)^{2}/\left(\det(A)\right)^{2}\\
 & =\det\left(AXA^{\top}\right)/\left(\det(A)\right)^{2}\\
 & =\left(\det\left(A\right)\det\left(X\right)\det\left(A\right)\right)/\left(\det(A)\right)^{2}\\
 & =\det\left(X\right).
\end{align*}
Therefore, either $p(X)=\textup{Pf}\left(X\right)$ for all $X$ or
$p(X)=-\textup{Pf}\left(X\right)$ for all $X$. Any matrix of positive determinant can be connect to $I$ by a path of matrices of positive determinant. (This fact follows from the singular value decomposition and the connectedness of real orthogonal matrices of determinant one.)
Thus $AXA^{\top}$ and $X$ are both in the same connected component
of $\textup{Sk}^{-1}(2n,\mathbb{R})$. Therefore $p(X)=\textup{Pf}\left(X\right)$
for all $X$, establishing the result in this case.

Now suppose $\det(A)$ is negative. Let $U$ be the matrix described
in Lemma~\ref{lem:rwo-column-swap-formula}. Then $UA$ has positive
determinant, so
\begin{align*}
\textup{Pf}\left(AXA^{\top}\right) & =\textup{Pf}\left(U\left(UAXA^{\top}U\right)U\right)\\
 & =-\textup{Pf}\left(UAX\left(UA\right)^{\top}\right)\\
 & =-\det(UA)\textup{Pf}\left(X\right)\\
 & =\det(A)\textup{Pf}\left(X\right).
\end{align*}

\vspace{5px}

We never establish a direct or recursive formula for the Pfaffian.
However, we can apply Theorem~\ref{thm:Det_Pfaff_of_AXA^T} repeatedly to compute a Pfaffian by simultaneously doing row and column operations.
More importantly, the sign of the Pfaffian can be calculated the same way. For the final step, we need to know the Pfaffian in the tridiagonal case.

\begin{lem}
\label{lem:Pfaffian_of_tridiagonal}If $T$ is a tridiagonal skew-symmetric
matrix in $\mathbf{M}_{2n}(\mathbb{C})$,
\begin{equation*}
T=\left[\begin{array}{cccccc}
0 & t_{1}\\
-t_{1} & 0 & t_{2}\\
 & -t_{3} & 0 & t_{3}\\
 &  & -t_{3} & 0 & \ddots\\
 &  &  & \ddots & \ddots & t_{2n-1}\\
 &  &  &  & -t_{2n-1} & 0
\end{array}\right]
\end{equation*}
 then
\begin{equation*}
\textup{Pf}(T)=\prod_{j=1}^{n}t_{2j-1}.
\end{equation*}
\end{lem}

\vspace{5px}
\noindent 
\textbf{Proof}
If $T$ is singular, this becomes trivial. Otherwise, it is just a
matter of setting the sign in 
\begin{equation*}
\textup{Pf}(T)=\pm\sqrt{\det(T)}=\pm\Biggl|\prod_{j=1}^{n}t_{2j-1}\Biggr|.
\end{equation*}
By the argument given in the proof of Theorem~\ref{thm:Path_to_a_standard_matrix}
tells us that $T$ is in $\textup{Sk}_{+}^{-1}(2n,\mathbb{R})$ if
an even number of the $t_{j}$ are negative and in $\textup{Sk}_{-}^{-1}(2n,\mathbb{R})$
otherwise. This pins down the sign.

\vspace{5px}

\section{Limitation of projection-based methods for local $\mathbf{Z}_2$-topology}
\label{sect_supp:limit_proj}

\begin{figure}[t]
\center
\includegraphics[width=\columnwidth]{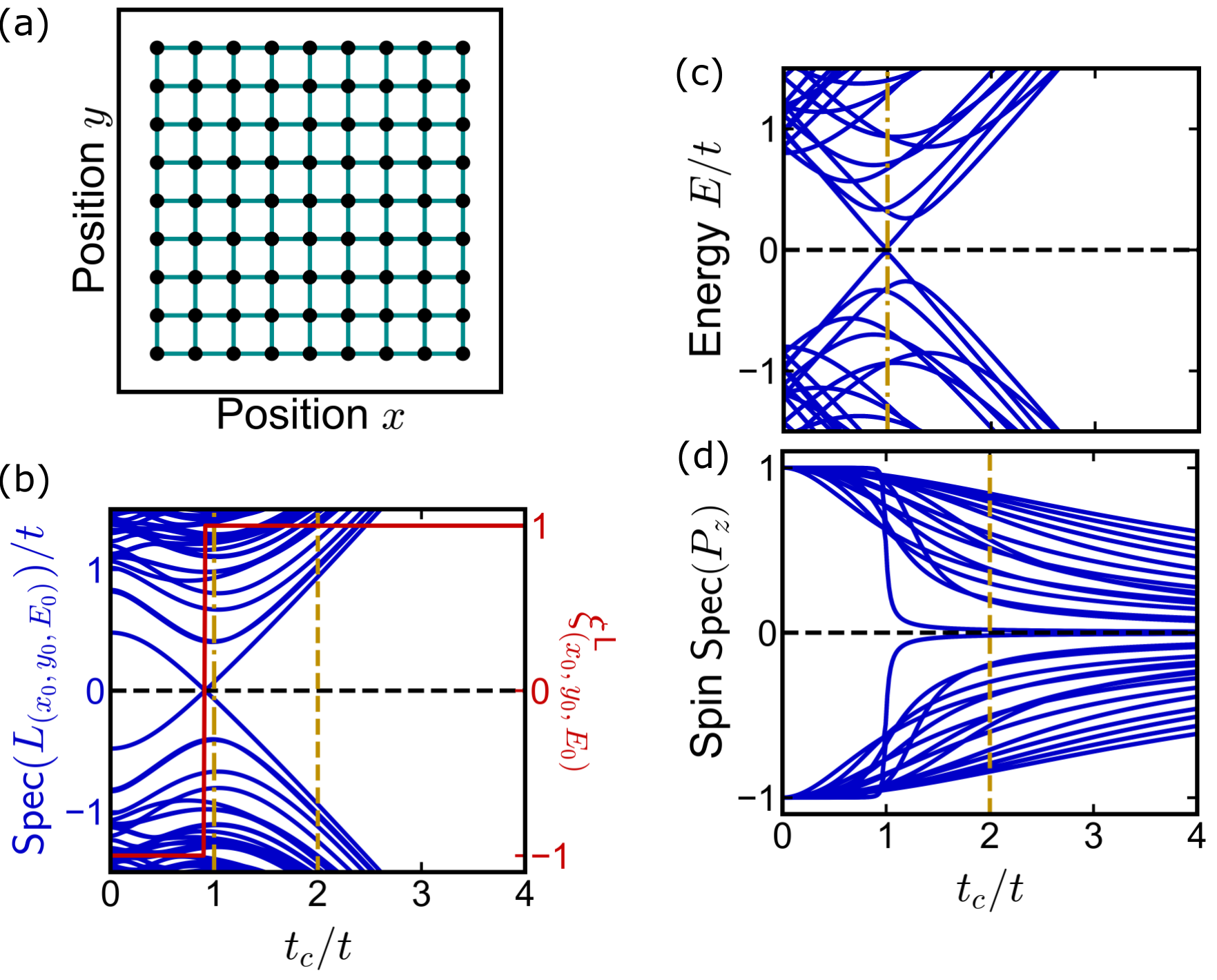}
\caption{
\textbf{Spectral gaps for local $\mathbb{Z}_2$-topology.}
(a) Square lattice for the Bernevig-Hughes-Zhang (BHZ) model, with parameters $t = 1$, $u = -1.2 t_1$, $t_1 = 0.5 t$ and $\Delta = t_1$.
(b) Spectrum of the spectral localizer $\text{Spec} \left( L_{(x_0,y_0,E_0)} \right)$ against the inter-layer couplings $t_c$, and the associated local $\mathbb{Z}_2$-topological invariant $\xi^\text{L}_{(x_0,y_0,E_0)}$ at the center of the lattice $(x_0,y_0)$ and energy $E_0=0$.
(c)-(d) Spectrum of the system's Hamiltonian $\text{Spec} \left( H \right)$ and the projected spin operator $\text{Spec} \left( P_z \right)$ against inter-layer couplings $t_c$.
The orange dotted-dashed and dashed vertical lines correspond to the spectral gap closing of the $H$ and $P_z$.
}
\label{fig_supp:bhz_spin}
\end{figure}

Compared to previous approaches to diagnose the local topology in non-periodic systems, the spectral localizer is not dependent on any spectral gaps in the Hamiltonian or projected operators such as the spin operator.

To illustrate the advantage of the spectral localizer, we consider the Bernevig-Hughes-Zhang (BHZ)~\cite{Bernevig2006} model which can be defined as two coupled square lattice layers 
\begin{equation}
\label{eq:H_qc_tr}
H = 
\left(
\begin{array}{cc}
H_1 & C \\
C^\dagger & H_1^*
\end{array}
\right)
,
\end{equation}
where $C = t_c I \otimes \sigma_y$ is the inter-layer coupling, acting on the internal degree of freedom, which can be seen as a spin-coupling, and $H_1$ and $H_1^*$ are the Hamiltonian associated to the first and second layer, respectively.
In particular, the Hamiltonian $H_1$ of the first square lattice layer has two internal degrees of freedom, where each vertex $j$ has a staggered on-site mass term
\begin{equation}
H_{1,jj} = u \sigma_z
,
\end{equation}
and the coupling between the vertices $j$ and $k$ is given by
\begin{equation}
H_{1,jk} = t_1 \sigma_z + i \frac{1}{2} \Delta \left( \sigma_x \cos(\theta_{jk}) + \sigma_y \sin(\theta_{jk}) \right)
,
\end{equation}
with $\theta_{jk} = 0, \pi/2, \pi, 3\pi/2$ the angle of the bond between the vertex $j$ and $k$ with respect to the (positive) horizontal axis, $t_1$ and $\Delta$ the couplings between the internal degree of freedom.
After, applying periodic boundary conditions to a finite size system such as in Fig.~\ref{fig_supp:bhz_spin}(a), the projected operator over the occupied states is defined as 
\begin{equation}
\label{eq_supp:proj}
P = \left[ \psi_1, \ldots, \psi_{N_\text{occ}} \right]
,
\end{equation}
where $\psi_i$ are the eigenvectors of the Hamiltonian $H$ associated with energy $E_i$ below the Hamiltonian's spectral gap, say at $E$, namely $H \psi_i = E_i \psi_i$ such that $E_i < E$.
The projected spin operator is then written
\begin{equation}
\label{eq_supp:proj_spin}
P_z = P \sigma_z P
,
\end{equation}
where $\sigma_z$ (the $z$-Pauli matrix) is the spin operator (in unit of $\hbar/2$).

For small enough spin-couplings $t_c$, there is a spectral gap for the projected spin operator $P_z$, namely the spin sectors remain well-defined with values around $-1$ and $1$, as shown in Fig.~\ref{fig_supp:bhz_spin}(d).
However, as the spin-mixing becomes stronger, the spin sectors start to merge, resulting in the closing of the spectrum of $P_z$.
As such, a $\mathbb{Z}_2$ invariant based on spin projectors or quantities derived from these projectors such as spin Chern numbers (e.g.\ the spin Bott index~\cite{Huang2018, Huang2018a}), are not well-defined in this regime and cannot be used to probe the system's topology.
Yet, the spectral localizer does not assume any spectral gaps but for the spectral localizer itself $L_{(\mathbf{x},E)}(X, Y, H)$, allowing it to probe material topology in regimes where the spectral gap in the Hamiltonian $H$ or in the projected spin operator $P_z$ closes, as demonstrated in Fig.~\ref{fig_supp:bhz_spin}(b).
Altogether, compared to the spectral localizer, projector-based approaches to local topology are not able to classify material topology over the full parameter space, as the spectral gap of the relevant operator might close.


\section{Chern topology in a quasicrystal}
\label{sect_supp:qc_chern}

\begin{figure}[t]
\center
\includegraphics[width=\columnwidth]{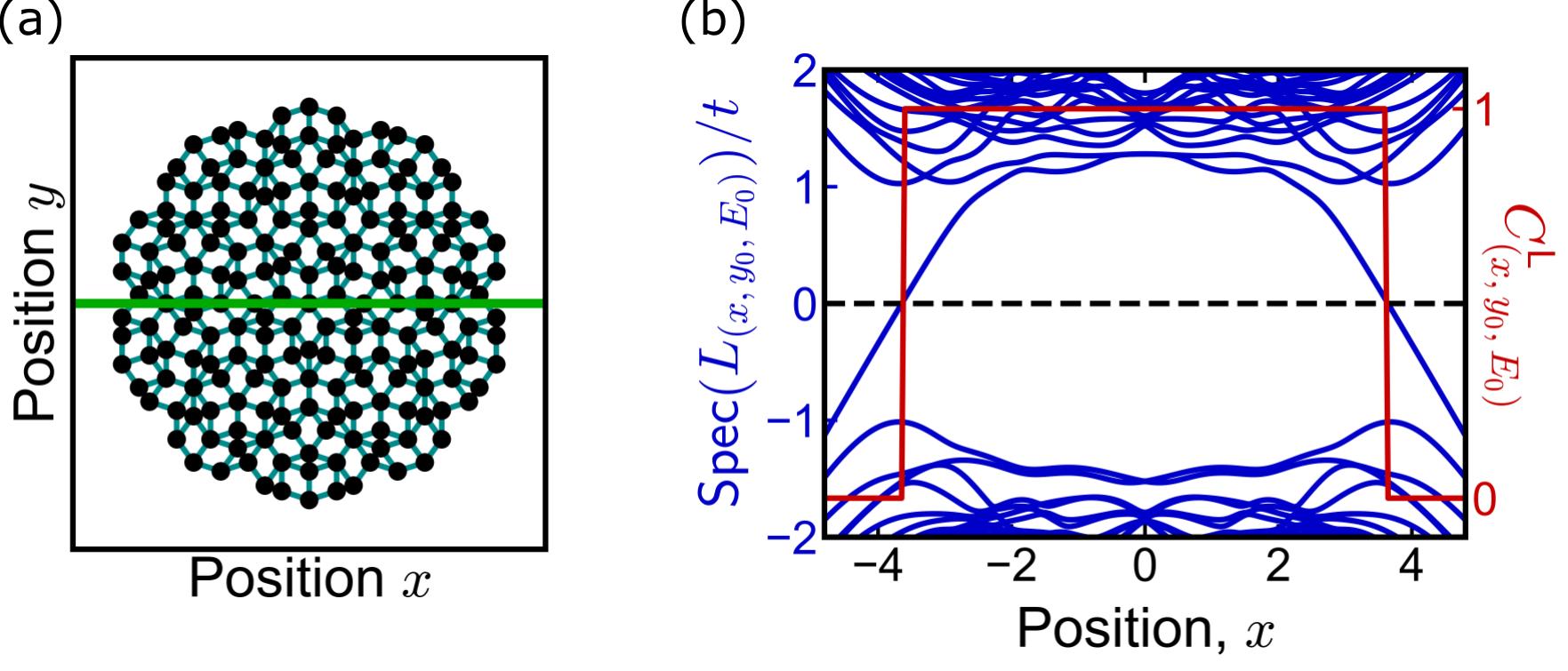}
\caption{
\textbf{Chern topology in quasicrystalline system.}
(a) Penrose tiling where the solid cyan lines represent the couplings between the internal degree of freedom on the vertices of the tiling, with parameters $t = 1$, $u = 2 t$, $t_1 = 2 t$ and $\Delta = 2 t$.
(b) Spectrum of the spectral localizer $\text{Spec} \left( L_{(x,y_0,E_0)} \right)$ along the solid green line and at energy $E_0=0$, and the associated local Chern-topological invariant $C^\text{L}_{(x,y_0,E_0)}$.
}
\label{fig_supp:qc_chern}
\end{figure}

Here, we demonstrate the non-trivial Chern topology for the single quasicrystalline layer considered in the main text, given by the Hamiltonian $H_1$ [Eqs.~\eqref{eq:H_qc_chern_jj}-\eqref{eq:H_qc_chern_jk}].

To diagnose the Chern topology in a two-dimensional (class A) system, the spectral localizer combines the system's Hamiltonian $H_1$ along with the system's position operators $X$ and $Y$ as the following~\cite{Loring2015}
\begin{align}
\label{eq_supp:localizer_chern}
\begin{split}
& L_{(\mathbf{x},E)}(X, Y, H_1) = \\[1.ex]
& 
\left(
\begin{array}{cc}
 (H - E I) & \kappa (X - x I)  - i \kappa (Y - y I) \\[1.ex] 
\kappa (X - x I) + i \kappa (Y - y I) & -(H - E I)
\end{array}
\right)
,
\end{split}
\end{align}
where $I$ is the identity matrix, 
and the associated local topological marker at specific spatial location $\boldsymbol{x} = (x,y)$ and energy $E$ is given by the local Chern number 
\begin{equation}
\label{eq:local_index}
C^\text{L}_{(\mathbf{x},E)}(X, Y, H_1) = 
\frac{1}{2} \text{sig} \left[ \tilde{L}_{(\mathbf{x},E)}(X, Y, H_1) \right]
\in \mathbb{Z}
,
\end{equation}
with $\text{sig}$ the signature spectral localizer.
Specifically $C^\text{L}_{(x,y,E)} \neq 0$ results in the case where the $X,Y,H$ matrices cannot be modified to be commuting, meaning the system is in the non-trivial topological phase locally at $(x,y,E)$.

Figure~\ref{fig_supp:qc_chern}(b) plots the spectral flow of the spectral localizer along the position $x$, thus demonstrating the the model used for the quasicrystal layer $H_1$ [Eqs.~\eqref{eq:H_qc_chern_jj}-\eqref{eq:H_qc_chern_jk}] has topological non-trivial Chern topology.

\end{document}